\documentclass[a4paper,twocolumn]{article}

\usepackage{url}			% Formatação com \url

\usepackage{graphicx}
\setkeys{Gin}{width=0.95\columnwidth}		% Tamanho das figuras
\begin{document}

\title{Google Earth Physics}

\author{C E Aguiar$^1$ and A R Souza$^{1,2}$ \\
$^1$ Instituto de F\'\i sica, Universidade Federal do Rio de Janeiro, Brasil \\
$^2$ Col\'egio Pedro II, Rio de Janeiro, Brasil}

\date{}

\maketitle	%Separa cabeçalho do texto.

\begin{abstract}
Google Earth photographs often show ships and their wakes in great detail. We discuss how the images can be used to calculate the velocity of these ships.
\end{abstract}

\section{Introduction}

Google Earth is a globe-imaging software, a ``geographic browser'', launched by Google in 2005~\cite{GE}. The program provides web access to a huge (Google-owned) database of satellite and aerial photographs covering the entire surface of Earth. Image resolution depends on how interesting the viewed point is, and ranges from 10~cm in some urban centres to 15~m in many rural areas~\cite{wikipedia}. Three-dimensional imagery is also available for several locations, with vertical resolution between 10~m and 90~m.
%Other names: geobrowser, virtual globe.

\begin{figure}[htb]
\begin{center}
\includegraphics{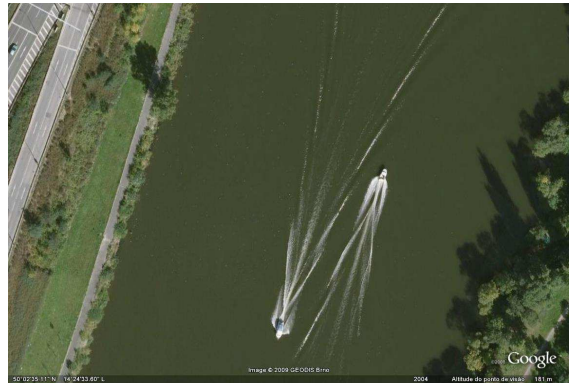}
\caption{Boat wakes on the Vltava river, Prague.}
\label{Prague} 
\end{center}
\end{figure}

Given the wealth of information available to Google Earth users, it is tempting to try the software in educational settings. The teaching of earth sciences is the most promising choice, but one could also think of applications to physics studies, as already pointed out in~\cite{Ryder}. We present here one such application: to determine the velocity of a ship from Google Earth pictures of its wake. This is possible because of the remarkable resolution of some imagery, which allows one to locate moving boats and observe details of their wakes. An example is in figure~\ref{Prague}, showing two boats and the wakes they create on the river Vltava, in Prague. As we will see, with simple kinematics and a basic knowledge of water waves, one can easily calculate the speed of boats such as those in figure~\ref{Prague} from analysis of their wakes.

\section{The Kelvin ship-wave pattern}

\begin{figure}[htb]
\begin{center}
\includegraphics{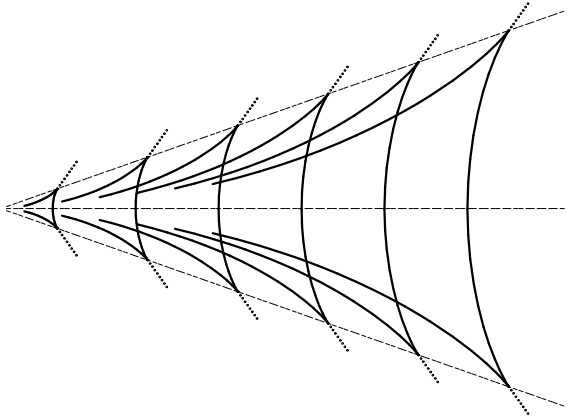}
\caption{Kelvin wave fronts on deep water. Dashed lines indicate the ship's path and the boundary of the Kelvin wedge. Evanescent waves extending beyond the wedge are shown as dotted lines.}
\label{kelvin} 
\end{center}
\end{figure}

Close to the vessel, the wake produced by a moving ship has a complex structure, which depends on the shape of the hull, the ship velocity and other factors. At large distances (compared to the dimensions of the hull) the situation becomes simpler and the wake takes a form with rather universal features -- the so-called Kelvin wave pattern~\cite{Kelvin}. The Kelvin waves follow the ship without changing shape (they appear stationary on the reference frame of the ship) and are localized in a wedge-like region behind the hull. 
Figure~\ref{kelvin} shows the Kelvin wave fronts for a ship on deep water. The ``Kelvin wedge'', bounded by lines making $\pm 19.5^\circ$ with the ship's path, contains most of the wake; only evanescent waves extend beyond this domain. Two kinds of waves can be noted inside the wedge: divergent waves, with V-shaped fronts that move away from the ship's path, and transverse waves that tend to follow the ship. Divergent and transverse waves form a cusp at the boundary of the Kelvin wedge; beyond that point the wave amplitude decays exponentially. A calculation of the Kelvin wave fronts based on geometric arguments can be found in~\cite{Lighthill}, together with a discussion of the many interesting properties of this wave pattern.

The amplitude along a wave front is not constant, being usually greatest close to the boundary of the Kelvin wedge. Also, divergent and transverse waves may have very different amplitudes. Small fast boats tend to produce large divergent waves and no transverse ones, as seen in figure~\ref{Prague}. Big slow ships, on the other hand, have their Kelvin pattern dominated by transverse waves. 

\section{Measuring ship speeds with Google Earth}

\begin{figure}[htb]
\begin{center}
\includegraphics[width=5cm]{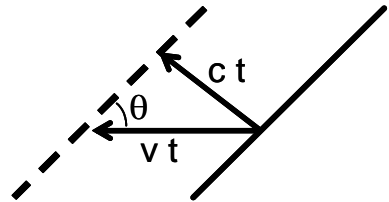}
\caption{Section of a wave front in the wake of a ship, shown at two different times separated by interval $t$. The ship moves from right to left with velocity $v$. The (phase) velocity of the wave is $c$ and the angle between the wave front and the ship's path is $\theta$.}
\label{kinematics} 
\end{center}
\end{figure}

For our purposes, the most important feature of the wake pattern is that it moves together with the ship. Because of this, there is a simple relation between the ship velocity $v$, the wave velocity $c$ and the angle $\theta$ the wave front makes with the ship's path:
\begin{equation}
v \sin \theta = c \;.
\label{vc}
\end{equation}
Equation~(\ref{vc}) is easily obtained from the kinematics of the wave pattern. This can be seen in figure~\ref{kinematics}, which shows part of a wave front at two instants separated by time $t$. During this time, the ship and the wave pattern move a distance $vt$, from right to left. On the other hand, the displacement of the wave in the direction perpendicular to the front is $ct$ ($c$ being the phase velocity). Simple trigonometry then leads to~(\ref{vc}). 

\begin{figure}[htb]
\begin{center}
\includegraphics{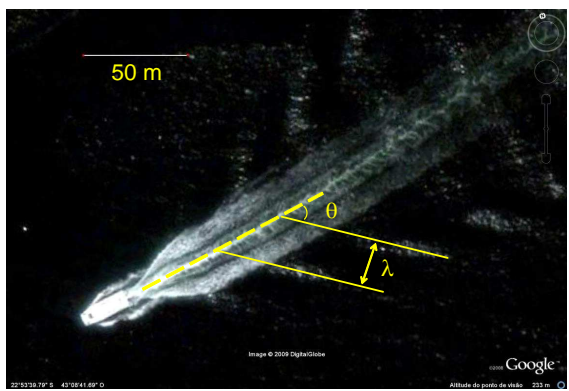}
\caption{Google Earth picture of a ferry boat near Rio de Janeiro. Measurements performed on the image give $\lambda=22 \,\mathrm{m}$ and $\theta = 43^\circ$. The $50 \,\mathrm{m}$ scale is fixed by Google Earth. The calculated boat speed is $v = 31 \,\mathrm{km/h}$.}
\label{PegasusRio} 
\end{center}
\end{figure}

On deep water, surface waves of wavelength $\lambda$ have phase velocity
\begin{equation}
c = \sqrt{\frac{g\lambda}{2\pi}}
\label{c}
\end{equation}
where $g$ is the acceleration of gravity~\cite{Crawford,Feynman}. Substituting this result into equation~(\ref{vc}), we obtain
\begin{equation}
v = \sqrt{\frac{g\lambda}{2\pi}}\,\frac{1}{\sin\theta} \;.
\label{v}
\end{equation}
The interesting thing about~(\ref{v}) is that it gives the ship speed as function of $\lambda$ and $\theta$, two quantities that can be measured directly from photographs such as those in Google Earth. Take for example the picture shown in figure~\ref{PegasusRio}, of a ferry boat heading towards the port of Rio de Janeiro. Using an image analysis program (\emph{ImageJ} is a good and free choice~\cite{ImageJ}), we measured the wavelength and angle indicated in the figure, obtaining $\lambda=22 \,\mathrm{m}$ and $\theta = 43^\circ$. With these results, and taking $g=9.8\,\mathrm{m/s}^2$, we find from~(\ref{v}) that the velocity of the ferry boat is $v = 31 \,\mathrm{km/h}$. This particular boat has been identified as catamaran \emph{P\'egasus}, which at the time the Google Earth picture was taken provided ferry service between the cities of Rio de Janeiro and Niter\'oi~\cite{private}. The cruise speed of \emph{P\'egasus} is $33 \,\mathrm{km/h}$~\cite{private}, quite close to the result of our calculation.

% $\lambda=22 \,\mathrm{m}\pm 2 \,\mathrm{m}$, $\theta = 43^\circ \pm 4^\circ$, 
% $L(boat) = 24 \,\mathrm{m}$, 
% $g=9.8\,\mathrm{m/s}^2$, $v=8.6 \,\mathrm{m/s} = 31 \,\mathrm{km/h}$.

\section{Conclusion}

We see from the above example that ship speeds can be measured easily and with good accuracy from Google Earth images. It is not even necessary to perform the analysis on a computer -- this may be carried out with ruler and compass on a printout of the image. If possible, one should look for pictures of boats used for public transportation in the school area, and ask the students to compare their calculations with a typical speed of the boat they analyzed. This limited search may not allow for the best choice of images, but it is always interesting to have the students check their results against other data (and it may be fun for them to take a boat ride just to ask the captain about the cruise speed). 

\section*{Acknowledgments}
This work was supported by Funda\c c\~ao de Amparo \`a Pesquisa do Estado do Rio de Janeiro (Faperj).

\end{document}